\documentclass[10pt, journal]{IEEEtran}
\usepackage{amsmath, amssymb, amsfonts}
\usepackage{graphicx}
\usepackage{epsfig}
\usepackage{subfigure}
\usepackage[font={small}]{caption}
\usepackage{cases}
\usepackage{tensor}

\usepackage{color}
\usepackage{epstopdf}
\usepackage{balance}

\usepackage{tikz}
\usepackage{cite}
\usepackage{tensor}
\usepackage{mathtools}

\makeatletter
\newcommand{\vast}{\bBigg@{4}}
\newcommand{\Vast}{\bBigg@{5}}
\makeatother

\begin{document}

\title{Massive MIMO versus Small-Cell Systems: Spectral and Energy Efficiency Comparison}
\author{ Hieu Duy Nguyen, \emph{Member, IEEE,} and Sumei Sun, \emph{Senior Member, IEEE}
\thanks{Manuscript received ...; revised ... ; accepted ... . This work was presented in part at ... . The associate editor coordinating the review of this paper and approving it for publication was ... .}
\thanks{The authors are with the Institute for Infocomm Research (I$^2$R), the Agency for Science, Technology and Research (A$\star$STAR), Singapore (email(s): \{nguyendh, sunsm\}@i2r.a-star.edu.sg).}
\thanks{Color versions of one or more of the figures in this paper are available online at http://ieeexplore.ieee.org.}
\thanks{Digital Object Identifier XXX.} \vspace{-0.1in}}

\maketitle


\newtheorem{definition}{\underline{Definition}}
\newtheorem{fact}{Fact}
\newtheorem{conjecture}{Conjecture}
\newtheorem{assumption}{\underline{Assumption}}
\newtheorem{theorem}{\underline{Theorem}}
\newtheorem{lemma}{\underline{Lemma}}
\newtheorem{corollary}{\underline{Corollary}}
\newtheorem{proposition}{\underline{Proposition}}
\newtheorem{example}{\underline{Example}}
\newtheorem{remark}{\underline{Remark}}
\newtheorem{algorithm}{\underline{Algorithm}}
\newcommand{\mv}[1]{\mbox{\boldmath{$ #1 $}}}

\captionsetup[table]{labelsep=newline}

\newcommand{\tikzcircle}[2][black,fill=black]{\tikz[baseline=-0.5ex]\draw[#1,radius=#2] (0,0) circle ;}%

\begin{abstract}
In this paper, we study the downlink performance of two important 5G network architectures, i.e. massive multiple-input multiple-output (M-MIMO) and small-cell densification. We propose a comparative modeling for the two systems, where the user and antenna/base station (BS) locations are distributed according to Poisson point processes (PPPs). We then leverage both the stochastic geometry results and large-system analytical tool to study the SIR distribution and the average Shannon and outage rates of each network. By comparing these results, we observe that for user-average spectral efficiency, small-cell densification is favorable in crowded areas with moderate to high user density and massive MIMO with low user density. However, small-cell systems outperform M-MIMO in all cases when the performance metric is the energy efficiency. The results of this paper are useful for the optimal design of practical 5G networks. 
\end{abstract}

\section{Introduction}\label{sec:intro}

The fifth generation (5G) communications system needs to deliver improvement over 4G, e.g. much higher user data rate/spectral efficiency, enhanced eco-friendliness and energy efficiency, seamless communications, and low latency etc. \cite{AnBuChHaLoSoZh14}. To meet the required performance, densified topologies with two main approaches, i.e., massive multiple-input multiple-output (M-MIMO) and small-cell densification, are promising candidate technologies \cite{JuMaZi14}.

In M-MIMO, each base station (BS) employs a large-scale antenna array in linear, cylindrical, or other shapes. The large number of antennas not only provides more diversity to the transmission but also ``hardens'' the channel, allowing low-complexity but sharp beamforming towards the user equipments (UEs) \cite{HoMaTa04}. The results are less interference and higher spectral and energy efficiencies. The main drawback of M-MIMO is the channel estimation/acquisition and feedback, which require a high accuracy to attain the gain that M-MIMO promises \cite{RuPeLaLaMaEdTu13,YiGeFiLi13,MaPi14}. Pilot contamination, itself an existing problem, also restricts the benefits of M-MIMO \cite{RuPeLaLaMaEdTu13,JoAsMaVi11}.

In contrast, small-cell networks consists of many micro/femto cells with limited number of antennas, typically one or two, at each access point (AP). Due to the short distance between the APs and UEs, small-cell networks have smaller path loss and co-channel interferences while requiring lower power consumption, hence improving both spectral and energy efficiencies. The gain of small-cell densification, however, is subject to more complicated cell planning and inter-cell coordination/cooperation \cite{HwSoSo13}.

The evolution of high spectral/energy efficiency for 5G thus results in two opposing ideas: concentrating the antennas to form M-MIMO BSs, and distributing them to form small cells. It is thus natural to ask which approach performs better and under which scenarios. Despite many works studying either M-MIMO or small-cell networks, there exists few work in literature comparing their performance. One of the reasons is due to system modeling: no existing framework so far allows for an effective collation. In this paper, we aim to partially answer the above question using the metrics of spectral and energy efficiencies. In particular, we resort to stochastic geometry and random matrix theory (RMT) for the analysis. A brief introduction of stochastic geometry can also be found in \cite{NgSu15}. 

RMT, also referred as large-system analysis in literature, was used to investigate large multi-user linear receivers and code division multiple access (CDMA) networks \cite{TsHa99,Ve98} and various MIMO setups \cite{Te99,FoGa98}. While exact analyses are realizable under finite-dimensional matrix theory \cite{Mc06}, they often lead to complicated expressions hence do not provide useful insights. The seminal monograph \cite{TuVe04} and other works have shown that large-system analysis can reveal many more interesting properties of communication systems. Furthermore, although the analysis is mathematically accurate only for large random vectors/matrices, the obtained results remain good approximations for finite-dimensional cases. Recently, many studies have also advanced the so-called ``deterministic equivalent'' approach to investigate more sophisticated system models \cite{CoDe11}.

M-MIMO, which employs a large number of transmit/receive antennas, is a natural application of large-system analysis. A detailed survey can be found in \cite{Ma15}, which lists many key M-MIMO performance analyses considering hardware impairments \cite{BjHoKoDe14}, channel state information (CSI) estimation/acquisition \cite{JoAsMaVi11,YiGeFiLi13}, and resource allocation/precoding schemes \cite{HuCaPaRa,NgLaMa13}. Note that most of the M-MIMO studies only consider a BS-centric analysis in which the relative locations of the BSs and UEs are fixed, thus neglecting the network topology. In order to compare M-MIMO with small-cell densification, a proper system modeling is necessary.

In this paper, we study both small-cell and M-MIMO networks using stochastic geometry and large-system tools. The aim is to provide an insightful comparison between these two topologies. The contributions of our paper are summarized as follows:
\begin{itemize}
\item \textbf{System Modeling:} We propose a model for M-MIMO and small-cell networks which incorporates the randomness of both transmitting distributed antennas (DAs)/BSs and single-antenna UEs. Particularly, the DA/BS and UE locations are distributed according to Poisson point process (PPPs) with possibly different densities. Therefore, there exists a transmission probability that some transmitting antennas/BSs are turned off due to no associated UE. We assume a time/frequency division multiplexing access (TDMA/FDMA) scheme in which the transmitting nodes associated with more than one UE will communicate with only one UE in each resource block. Finally, we assume non-coordinated DAs/BSs and employ conjugate beamforming, which is optimal for non-coordinated M-MIMO with single-antenna UEs.
\item \textbf{M-MIMO and Small-Cell Network Analysis:} We study the performance of M-MIMO and show that the signal-to-interference ratio (SIR) distribution at a typical UE is similar to that under a non-fading ad-hoc network. Here, the important differences between the two cases are the M-MIMO gain and transmission probability. As discussed before, the results for fading ad hoc networks can be directly extended to the small-cell counterparts, with small but important modification again due to the transmission probability. We therefore utilize the results from \cite[Sections III and IV]{NgSu15} to derive the bounds for the SIR distribution of M-MIMO and small-cell networks. Based on these analyses, we further obtain the bounds for both the average Shannon and outage rates of the two systems. To better characterize the small interference regime, the performance of M-MIMO and small-cell networks is also derived under the assumption of asymptotically small UE density.
\item \textbf{Comparison between M-MIMO and Small-Cell Densification:} We reveal that small-cell densification yields better rates than M-MIMO when the user density is moderate or large compared to the transmitting node density and number of antennas. However, M-MIMO outperforms small-cell systems under the asymptotically small UE density regime. This implies that there exists a UE density threshold lower than which we should employ M-MIMO and higher than which small-cell densification is more preferable, when spectral efficiency is the performance metric. We further compare the energy efficiency performance and show that small-cell always outperforms M-MIMO. The results are therefore useful for the optimal design of the upcoming 5G heterogeneous networks.  
\end{itemize}

The rest of the paper is organized as follows. Section \ref{sec:system model} describes our M-MIMO and small-cell network models, which incorporate the randomness of both transmitting nodes and UEs, for a fair comparison between M-MIMO and small-cell densification. In Sections \ref{sec:M-MIMO CAS} and \ref{sec:small-cell}, we investigate the M-MIMO and small-cell systems introduced in Section \ref{sec:system model}, based on the results from \cite[Sections III and IV]{NgSu15}. The spectral and energy efficiency comparisons between M-MIMO and small-cell densification are given in Sections \ref{sec:comparison} and \ref{sec:comparison EE}, respectively. Finally, Section \ref{sec:conclusions} concludes the paper with some remarks.

{\it Notations}: Scalars and vectors/matrices are denoted by lower-case and bold-face lower-case/upper-case letters, respectively. The conjugate, transpose, and conjugate transpose operators are denoted by $(\cdot)^*$, $(\cdot)^T$, and $(\cdot)^H$, respectively. $[\mv{A}]_{i,j}$ stands for the $(i,j)$th element of the matrix $\mv{A}$. $\mathbb{E}_{X}[\cdot]$ denotes the statistical expectation over a random variable $X$. $\mathrm{Tr}(\cdot)$, $||\cdot||_F$, and $\det (\cdot)$ represent the trace, Frobenius norm, and determinant of a matrix, respectively. $\mv{I}_{M}$ denotes an $M$-by-$M$ identity matrix. $\mathbb{R}^{x \times y}$ and $\mathbb{C}^{x \times y}$ denote the space of $x$-by-$y$ real and complex matrices, respectively. Finally, the circularly symmetric complex Gaussian distribution with mean $\mu$ and variance $\sigma^2$ is represented by $\mathcal{CN}(\mu,\sigma^2)$.

\section{Stochastic Modeling for Massive MIMO and Small-Cell Systems}\label{sec:system model}

\subsection{M-MIMO System}
We consider a M-MIMO system where BS and UE locations are distributed according to homogeneous PPPs $\Phi_{C,d}$ and $\Phi_{C,u}$ of intensities $\lambda_{C,b} = \lambda_b$ and $\lambda_{C,u} = \lambda_{D,u} = \lambda_{u}$, respectively. Each UE is associated with the nearest BS each with $M$ antennas. Here, we assume that each BS is scheduled to serve only one UE at each time/frequency slot due to time/frequency division multiple access (TDMA/FDMA). In each slot, the received signal of the UE associated with BS $b$ is given as
\begin{align}
y_{C,b} = \mv{h}_{C,b,b}^H \mv{x}_{C,b} + \displaystyle \sum_{ b'\in \mathcal{B} /\{ b \} } \mv{h}_{C,b',b}^H \mv{x}_{C,b'} + n_{C,b},
\end{align}
where $\mathcal{B}$ is the set of transmitting BSs; $\mv{h}_{C,b',b}^H$ is the channel between BS $b'$ and the associated UE of BS $b$; and $\mv{x}_{C,b'}$ is the transmit signal from BS $b'$. Here, $\mv{h}_{C,b',b} = \sqrt{\xi_{C,b',b}} \widehat{\mv{h}}_{C,b',b}$; $\xi_{C,b',b}$ = $d_{C,b',b}^{-\mu}$ is a large scale fading including path loss with $\mu$ denoting the path loss exponent; $\widehat{h}_{C,b',b}$ follows a statistical distribution $\widehat{h}_{C,b',b} \sim \mathcal{CN}(0,1),~\forall b, b' \in \mathcal{B}$; and $n_{C,b}$ is the additive white Gaussian noise (AWGN) distributed as $\mathcal{CN}(0,\sigma_n^2)$. For a fair comparison with small-cell systems, we assume that each BS has a total power constraint of $MP_T$ and therefore $\mathbb{E}|\mv{x}_{C,b'}|^2 = M P_T$.

We consider a practical regime in which each BS only knows the channel between itself and the associated UE via, e.g., channel estimation and feedback. Under such assumption, each BS implements the conjugate beamforming to transmit signals to its respective UEs. 

Fig. \ref{fig:MMIMO ChRel} shows one channel realization of the M-MIMO system model. Here, the service area is assumed to be a square region of $20\times 20$. The BS and user densities are $\lambda_{C,b} = 0.05$ and $\lambda_u = 0.15$, respectively. Note that as the UE density is not sufficiently larger than the transmitter density, some BSs are not associated to any user. The corresponding cells are marked by green color for a clear illustration.

\subsection{Small-Cell System}

\begin{figure}[t]
 		\centering 
 		\epsfxsize=0.48\linewidth
 		\captionsetup{width=0.48\textwidth} 
 		\includegraphics[width=6.5cm, height=6.5cm]{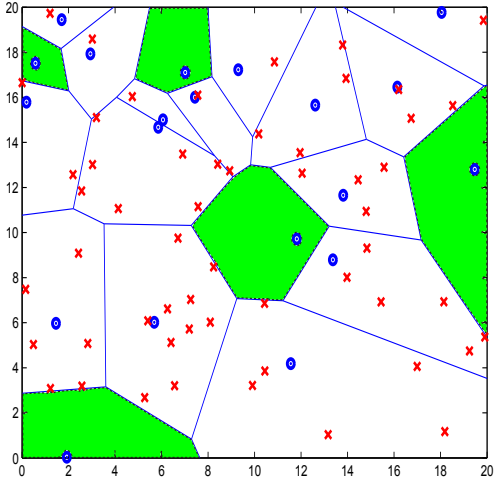}
 		\caption{A channel realization for either M-MIMO or small-cell systems. Here ``o'', ``x'', and straight lines denote the transmitter (BS/DA), user, and boundaries between cells. The cells without any user are marked by green color.}\label{fig:MMIMO ChRel}
\vspace{-0.1in}
\end{figure}

For a fair comparison, we consider a small-cell system where distributed antenna (DA) and UE locations are distributed according to homogeneous PPPs $\Phi_{D,b}$ and $\Phi_{D,u}$ of intensities $\lambda_{D,b} = M \lambda_b$ and $\lambda_{D,u} = \lambda_{u}$, respectively, in the Euclidean plane. Each DA is associated with the nearest UE. For simple description, we consider one resource block (e.g., frequency and time), which is taken by only one UE, i.e., one DA serves one user at a time. The received signal at user $k$ is
\begin{align}
y_{D,k} = h_{D,k,k}x_{D,k} + \sum_{k'\in \mathcal{K} / \{k\} } h_{D,k,k'} x_{D,k'} + n_{D,k},
\end{align}
where $\mathcal{K}$ is the user set, $h_{D,k,k'}$ is channel gain between user $k$ and selected DA for user $k'$; $x_{D,k}$ is the transmit signal from user $k$; and $n_{D,k}$ is the AWGN $\mathcal{CN}(0,\sigma_n^2)$. Here, $h_{D,k,k'} = \sqrt{\xi_{D,k,k'}} \widehat{h}_{D,k,k'}$; $\xi_{D,k,k'}$ = $d_{D,k,k'}^{-\mu}$ is a large scale fading including path loss with $\mu$ denoting the path loss exponent; and $\widehat{h}_{D,k,k'}$ follows a statistical distribution $\widehat{h}_{D,k,k'}\sim \mathcal{CN}(0,1),~\forall k,k'\in\mathcal{K},k' \neq k$. We assume equal power constraint at each DA, i.e., $\mathbb{E}|x_{D,k}|^2 = P_T$. Note that we can consider Fig. \ref{fig:MMIMO ChRel} as a channel realization for the small-cell system model with the DA and UE densities being $\lambda_{D,b} = 0.05$ and $\lambda_u = 0.15$.

\section{Massive MIMO System Analysis}\label{sec:M-MIMO CAS}

In this section, we investigate the performance of M-MIMO systems based on the results in our paper \cite[Section II]{NgSu15}. First, we derive the distribution of the signal-to-interference-plus-noise ratio (SINR) and SIR. Note that if the UE intensity $\lambda_u$ is not sufficiently larger than $\lambda_b$, then some M-MIMO BSs may not possibly be associated to any UE, and thus, do not transmit signals. Therefore, it is necessary to study the transmission probability of each BS. The following result provides the area distribution of Voronoi cells.
\begin{lemma}[\cite{FeNe07}]\label{lemma:area PPP}
Given a PPP $\Phi$ with homogeneous density $\lambda$. The probability distribution function (PDF) of $S$, the area of a typical Voronoi cell formed from $\Phi$, can be approximated as 
\begin{align}
f_S(x) \approx \frac{3.5^{3.5}}{\Gamma(3.5)} \lambda^{3.5} x^{2.5} e^{-3.5 \lambda x}.
\end{align}
\end{lemma}

By using Lemma \ref{lemma:area PPP}, the transmission probability of BSs is approximated as \cite{LeHu12}
\begin{align}\label{eq:epsC}
\epsilon_C \approx 1 - \left( 1+ \frac{\lambda_u}{3.5\lambda_b} \right)^{-3.5}.
\end{align}

Note that given the transmission probability $\epsilon_C$, the transmitting BS locations are thus distributed according to a PPP $\widehat{\Phi}_{C,b}$ with density $\widehat{\lambda}_{C,b}$ $= \epsilon_C \lambda_b$.

\subsection{Conjugate Beamforming - Large System Analysis}

Since all BSs employ conjugate beamforming, the transmit signal $\mv{x}_{C,b}$ is
\begin{align}
\mv{x}_{C,b} = \mv{w}_{C,b} s_{C,b} = \frac{\sqrt{MP_T} \mv{h}_{C,b,b}}{\lVert \mv{h}_{C,b,b} \rVert} s_{C,b},
\end{align}
where $\mv{w}_{C,b}$ $=$ $\sqrt{MP_T} \mv{h}_{C,b,b} / \lVert \mv{h}_{C,b,b} \rVert$ $\in \mathbb{C}^{M \times 1}$ is the conjugate beamformer of BS $b$. As a consequence, the received signal of the UE associated to BS $b$ is given as follows
\begin{align}
y_{C,b} 
& = \sqrt{MP_T} \lVert \mv{h}_{C,b,b} \rVert s_{C,b} \notag \\
& + \displaystyle \sum_{ b'\in \mathcal{B} /\{ b \} }  \frac{\sqrt{MP_T} \mv{h}_{C,b',b}^H \mv{h}_{C,b',b'}}{\lVert \mv{h}_{C,b',b'} \rVert} \mv{x}_{C,b'} + n_{C,b},
\end{align}

The SINR of the UE associated to BS $b$ therefore can be expressed as
\begin{align}\label{eq:SINR CAS}
\text{SINR}_{b} 
& = \frac{ MP_T  \lVert \mv{h}_{C,b,b} \rVert^2 } { \sum\limits_{ b'\in \mathcal{B} /\{ b \} } \displaystyle MP_T \frac{ \mv{h}_{C,b',b}^H \mv{h}_{C,b',b'} \mv{h}_{C,b',b'}^H \mv{h}_{C,b',b} }{ \lVert \mv{h}_{C,b',b'} \rVert^2 } + \sigma_n^2} \notag \\
& = \frac{ d_{C,b,b}^{-\mu}  \left\lVert \widehat{\mv{h}}_{C,b,b} \right\rVert^2 } { \sum\limits_{ b'\in \mathcal{B} /\{ b \} } \displaystyle d_{C,b',b}^{-\mu} \frac{ \widehat{\mv{h}}_{C,b',b}^H \widehat{\mv{h}}_{C,b',b'} \widehat{\mv{h}}_{C,b',b'}^H \widehat{\mv{h}}_{C,b',b} }{ \left\lVert \widehat{\mv{h}}_{C,b',b'} \right\rVert^2 } + \frac{\sigma_n^2}{MP_T}}.
\end{align}

Note that the vector series $\frac{\widehat{\mv{h}}_{C,b,b}} {\sqrt{M}}$ $\in \mathbb{C}^{M\times 1}$ satisfy the condition of \cite[Theorem 3.4]{CoDe11}. Therefore, we have
\begin{align}\label{eq:rmt 1}
\left\lVert \widehat{\mv{h}}_{C,b,b} \right\rVert^2/M \xrightarrow{M\to \infty} 1,
\end{align}

Based on (\ref{eq:rmt 1}), the SINR of the UE associated to BS $b$ is expressed as
\begin{align}\label{eq:SINR Pre Asymp}
& Q_{\text{M}} \triangleq \text{SINR}_{b} \notag \\
&= \frac{ d_{C,b,b}^{-\mu}  \left\lVert \widehat{\mv{h}}_{C,b,b} \right\rVert^2/M } { \sum\limits_{ b'\in \mathcal{B} /\{ b \} } \displaystyle d_{C,b',b}^{-\mu} \frac{ \widehat{\mv{h}}_{C,b',b}^H \widehat{\mv{h}}_{C,b',b'} \widehat{\mv{h}}_{C,b',b'}^H \widehat{\mv{h}}_{C,b',b} }{ M \left\lVert \widehat{\mv{h}}_{C,b',b'} \right\rVert^2 } + \frac{\sigma_n^2}{M^2P_T}} \notag \\
& \xrightarrow{M\to \infty} \frac{ M d_{C,b,b}^{-\mu}  } { \sum\limits_{ b'\in \widehat{\Phi}_{C,d} /\{ b \} } \displaystyle d_{C,b',b}^{-\mu} |\tilde{h}_{b'}|^2 + \frac{\sigma_n^2}{MP_T}},
\end{align}
where $ \tilde{h}_{b'} \triangleq \frac{ \widehat{\mv{h}}_{C, b',b}^H \widehat{\mv{h}}_{C, b',b'} }{ \left\lVert \widehat{\mv{h}}_{C, b',b'} \right\rVert }$; $\tilde{h}_{b'}$'s are i.i.d. random variables each distributing as $\mathcal{CN}(0,1)$. The form (\ref{eq:SINR Pre Asymp}) is similar to that investigated in \cite[Section V]{NgSu15}. It has been shown that (\ref{eq:SINR Pre Asymp}) is difficult to study, in particular to derive the cumulative distribution probability (CDF) bounds. As discussed in \cite[Section V]{NgSu15}, we can well approximate
\begin{align}\label{eq:INR approx}
\sum\limits_{ b'\in \widehat{\Phi}_{C,d} /\{ b \} } \displaystyle d_{C,b',b}^{-\mu} |\tilde{h}_{b'}|^2 \approx \sum\limits_{ b'\in \widehat{\Phi}_{C,d} /\{ b \} } \displaystyle d_{C,b',b}^{-\mu}.
\end{align}

As a consequence, we have 
\begin{align}\label{eq:SINR Asymp}
& Q_{\text{M}} \approx \frac{ M d_{C,b,b}^{-\mu}  } { \sum\limits_{ b'\in \widehat{\Phi}_{C,d} /\{ b \} } \displaystyle d_{C,b',b}^{-\mu} + \frac{\sigma_n^2}{MP_T}}.
\end{align}

\subsection{Probability Distribution}  

We note that the form of the SINR in (\ref{eq:SINR Asymp}) is similar to that investigated in \cite[Section II]{NgSu15}. Therefore, we can apply the results for the outage probability, average achievable rate, and outage rate in \cite[Section II]{NgSu15} here. The only but important difference is that the interfering BSs' locations are distributed according to the PPP $\widehat{\Phi}_{C,d}$ with density $\widehat{\lambda}_d$, while the PDF of the nearest distance follows that from PPP $\Phi_{C,d}$ with density $\lambda_b$. The following corollaries follow same reasonings as in \cite[Theorem 1, (9), and (10)]{NgSu15}. The proofs are given in Appendix \ref{appen:MMIMO CDF}, where we only highlight the differences for brevity. 
\begin{corollary}\label{corol:MMIMO CDF SINR}
Assume that the transmitting BSs follows a PPP $\widehat{\Phi}_{C,d}$ with density $\widehat{\lambda}_{C,d}$. The Laplace transform of $Q_{\text{M}}^{-1}$ can be expressed as
\begin{align}
& \mathcal{L}_{Q_{\text{M}}^{-1}} \left(s \right) 
= \pi \lambda_b \int_{0}^{\infty}  \exp \bigg\{ -\frac{s y^{\mu/2} \sigma_n^2}{M^2 P_T} - \pi \widehat{\lambda}_{C,d} y \notag \\
& \left[ 1 + \frac{1}{\epsilon_C} - e^{-s/M} + \left( \frac{s}{M} \right)^{2/\mu} \gamma\left( 1- \frac{2}{\mu}, \frac{s}{M} \right)  \right]   \bigg\} dy,
\end{align}
\end{corollary}

\begin{corollary}\label{corol:MMIMO CDF bounds}
Given that $M\to\infty$, the lower and upper bounds of the SIR distribution are given as follows
\begin{subnumcases}{\label{eq:MMIMO LB all Q} F^{\text{LB}}_{Q_\text{M}}(q) = }
0, &  $0\leq q < M$, \label{case:MMIMO LBqLeq1} \\
1 - \frac{1}{\epsilon_C M^{-\frac{2}{\mu}} q^{\frac{2}{\mu}} + 1 - \epsilon_C}, \hspace{-0.2in} &  $M \leq q$, \qquad \quad \label{case:MMIMO LBqGeq1}
\end{subnumcases} 

\begin{flalign}
~~ F^{\text{UB}}_{Q_\text{M}}(q) & = 1 - \frac{1}{1 + \epsilon_C \beta_{\text{n-fd}} M^{-2/\mu} q^{2/\mu}}. & \label{eq:MMIMO UB all Q}
\end{flalign} 
\end{corollary}

Similar to \cite[(10)]{NgSu15}, (\ref{eq:MMIMO UB all Q}) is rigorously proved only for: (a) $q \geq M$, and (b) $q<M$ with $q\to 0$ or $q\to M$. The following result is a counterpart of \cite[Corollary 1]{NgSu15} for massive MIMO when $M\to\infty$. The proof can be deduced from Appendix \ref{appen:MMIMO CDF bounds}, and thus is omitted for brevity. Note that Corollary \ref{corol:E SINR^-1} does not require the approximation (\ref{eq:INR approx}) since $\mathbb{E}_{\Phi} \left[ M \text{SIR}_{b}^{-1} \right]$ already averages over the channel fading.

\begin{corollary}\label{corol:E SINR^-1}
Assuming a very large $M$, the expectation of $M \text{SIR}_{b}^{-1}$ is given as $\mathbb{E}_{\Phi} \left[ M \text{SIR}_{b}^{-1} \right] = \frac{2 \epsilon_C}{\mu-2}$. 
\end{corollary}

\begin{remark}
It is important to remind that Corollaries \ref{corol:MMIMO CDF bounds} and \ref{corol:E SINR^-1} are derived by assuming that the noise power is negligible, i.e., $\sigma^2 = 0$, and considering the SIR instead of SINR. In small UE density network, i.e., $\lambda_b \gg \lambda_u$, such assumption is not accurate anymore since the interfering sources are far away and the interfering power is small. The noise effect therefore is more important. As a consequence, Corollaries \ref{corol:MMIMO CDF bounds} and \ref{corol:E SINR^-1} are not accurate in UE-sparse networks. Further results for M-MIMO with asymptotically small UE density will be presented in Section \ref{subsec:asymp small UE den}.
\end{remark}

\begin{figure}[t]
 		\centering 
 		\epsfxsize=0.48\linewidth
 		\captionsetup{width=0.48\textwidth} 
 		\includegraphics[width=8.5cm, height=6.5cm]{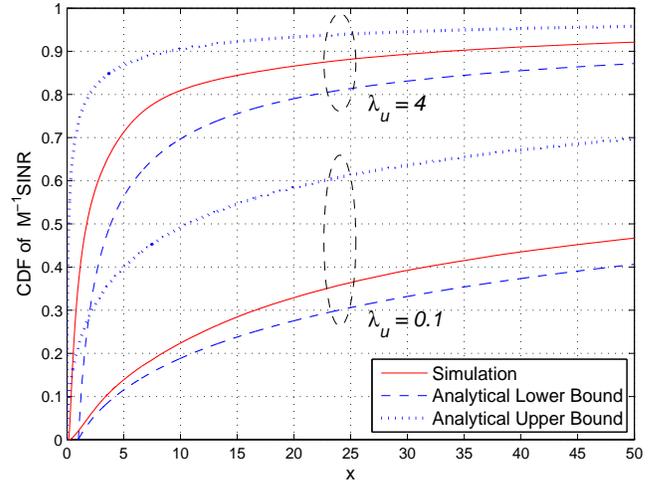}
 		\caption{Simulated SINR CDF and its bounds (\ref{eq:MMIMO LB all Q}), (\ref{eq:MMIMO UB all Q}) for M-MIMO with $M = 64$, $\mu = 3.7$, $\lambda_b = 1$, and $P_T/\sigma^2 = 15$ dB.}\label{fig:CDF MMIMO}
\vspace{-0.1in}
\end{figure}

\begin{figure}[t]
 		\centering 
 		\epsfxsize=0.48\linewidth
 		\captionsetup{width=0.48\textwidth} 
 		\includegraphics[width=8.5cm, height=6.5cm]{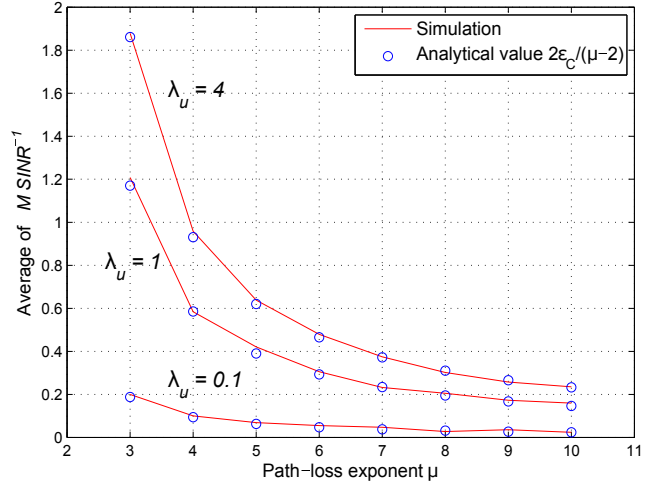}
 		\caption{Average of $M \text{SINR}_{b}^{-1}$ for M-MIMO with $M = 64$, $\mu = 3.7$, $\lambda_b = 1$, and $P_T/\sigma^2 = 15$ dB.}\label{fig:SINRInv MMIMO}
\vspace{-0.1in}
\end{figure}

In Figs. \ref{fig:CDF MMIMO} and \ref{fig:SINRInv MMIMO}, we confirm the validity of Corollaries \ref{corol:MMIMO CDF bounds} and \ref{corol:E SINR^-1}, respectively. From Fig. \ref{fig:SINRInv MMIMO}, it is observed that the assumption that the transmitting BSs follow a PPP $\widehat{\Phi}_{C,d}$ with density $\widehat{\lambda}_{C,d} = \epsilon_C \lambda_b$ with $\epsilon_C$ given in (\ref{eq:epsC}) is quite accurate. However, the upper bound is only close to the exact SINR distribution when $\lambda_u$ is sufficiently large ($\lambda_u \geq \lambda_b$ by intensive simulations), as shown in Fig. \ref{fig:CDF MMIMO}. As $\lambda_u$ decreases compared to $\lambda_b$, the upper bound (\ref{eq:MMIMO UB all Q}) becomes looser. The reason is that some inequalities, used to derive the upper bound in Appendix \ref{appen:MMIMO CDF bounds}, are not strict especially when $\epsilon_C$ is small. The lower bound, however, is not affected by the UE density due to its simple derivation.

\subsection{Average Achievable and Outage Rate}

Based on Corollary \ref{corol:MMIMO CDF bounds}, we investigate the average Shannon and outage rates for M-MIMO systems. We note that due to TDMA/FDMA, each user is only served for a fraction of time or a sub-band. Assuming that all users have the same transmission priority, that fraction can be defined as $\frac{1}{N_{b,u}}$, where $N_{b,u}$ is the number of users in cell $b$. The following result establishes $\mathbb{E} \left[ N_{b,u} \right]$.

\begin{lemma}\label{lemma:Avg N_bu}
The expectation of the number of users in a typical cell $b$ is given as 
\begin{align}
\mathbb{E} \left[ N_{b,u} \right] = \frac{\lambda_{C,u}}{\lambda_{C,b}} = \frac{\lambda_u}{\lambda_b}. 
\end{align}
\end{lemma}
\begin{IEEEproof}
Please refer to Appendix \ref{appen:Avg N_bu}.
\end{IEEEproof}

As expected, the average number of users in a cell is proportional to the UE density while inversely proportional to the BS density. The simple linear scaling is surprising, but intuitive. The reason is due to the homogeneity of both BS and UE PPPs, which does not create clusters of either BSs or UEs. Therefore, with $\lambda_u$ UEs and $\lambda_b$ BSs in a region of $1$ km$^2$, we expect to see the UEs homogeneously distributed across all BSs, which leads to $\lambda_u/\lambda_b$. By comparing the simulated and analytical values of $\mathbb{E} \left[ N_{b,u} \right]$, intensive simulations confirm the validity of Lemma \ref{lemma:Avg N_bu}. The comparison is not presented due to the space constraint.

Note that the rate of an UE associated to BS $b$ is given as
\begin{align}
R_{b,u} = \frac{1}{N_{b,u}} \log_2 \left( 1 + \text{SINR}_b \right),
\end{align}
where the fraction $\frac{1}{N_{b,u}}$ is due to TDMA/FDMA. We thus obtain the average Shannon rate of a typical user as follows
\begin{align}
R_{\text{M}} 
= \mathbb{E}_{\Phi_{C,b},\Phi_{C,u}} \left[ R_{b,u} \right] 
\overset{(a)}{=}  \mathbb{E} \left[ \frac{1}{N_{b,u}} \right] \mathbb{E} \left[ \log_2 \left( 1 + \text{SINR}_b \right) \right].
\end{align}
Here, $(a)$ is due to TDMA/FDMA and the fact that the random variables $\text{SINR}_b$ and $N_{b,u}$ are independent given that TDMA/FDMA is applied.

Based on Corollary \ref{corol:MMIMO CDF bounds} and Lemma \ref{lemma:Avg N_bu}, we can derive the bounds for the average achievable rate and the outage rate as follows. The proofs are similar to that of \cite[Lemmas 3 and 4]{NgSu15} and omitted for brevity. 
\begin{corollary}\label{corol:MMIMO Ravg UB LB}
The average user rate $R_{\text{M}}$ is bounded above and below by $R^{\text{UB}}_{\text{M}}$ and $R^{\text{LB}}_{\text{M}}$, respectively, where
\begin{align}\label{eq:MMIMO R UB}
R^{\text{UB}}_{\text{M}} 
& = \min \left\{ 1, \frac{\lambda_b}{\lambda_u} \right\} \Bigg( \log_2(1+M) \notag \\
& + \int_{\log_2(1+M)}^{\infty} \frac{dt}{\epsilon_C M^{-2/\mu} (2^t-1)^{2/\mu} + 1 - \epsilon_C } \Bigg),
\end{align}
\begin{align}\label{eq:MMIMO R LB}
R^{\text{LB}}_{\text{M}} = \min \left\{ 1, \frac{\lambda_b}{\lambda_u} \right\}  \int_{0}^{\infty} \frac{dt}{ 1 + \epsilon_C \beta_{\text{n-fd}} M^{-2/\mu} (2^t-1)^{2/\mu} }.
\end{align}
\end{corollary}

\begin{corollary}\label{corol:MMIMO OR UB LB}
The outage user rate $\text{OR}_{\text{M}}(\eta)$ is upper- and lower-bounded by $\text{OR}_{\text{M}}^{\text{UB}}(\eta)$ and $\text{OR}_{\text{M}}^{\text{LB}}(\eta)$, respectively, where \\
$\text{OR}_{\text{M}}^{\text{UB}}(\eta)$
\begin{subnumcases}{\label{eq:MMIMO OR UB} =}
\min \left\{ 1, \frac{\lambda_b}{\lambda_u} \right\} \log_2(1+\eta), & $0 \leq \eta \leq M$, \qquad \label{case:MMIMO OR UB etaSmall} \\
\min \left\{ 1, \frac{\lambda_b}{\lambda_u} \right\} \frac{\log_2(1+\eta)}{\epsilon_C \left( \frac{\eta}{M} \right)^{2/\mu} + 1 - \epsilon_C} , \hspace{-0.2in} & $M < \eta$. \label{case:MMIMO OR UB etaLarge}
\end{subnumcases}
\begin{flalign}\label{eq:MMIMO OR LB}
\text{OR}^{\text{LB}}_{\text{M}}(\eta) 
& = \min \left\{ 1, \frac{\lambda_b}{\lambda_u} \right\} \frac{\log_2 \left(1+\eta \right)}{1 + \epsilon_C \beta_{\text{n-fd}} \left( \frac{\eta}{M} \right)^{2/\mu} }. &
\end{flalign}
\end{corollary}

Here, the term $\min \left\{ 1, \frac{\lambda_b}{\lambda_u} \right\}$ in Corollaries \ref{corol:MMIMO Ravg UB LB} and \ref{corol:MMIMO OR UB LB} is due to the fact that TDMA is applied only when $N_{b,u} > 1$, and we have approximated 
\begin{align}\label{eq:approx 1/N}
\mathbb{E} \left[ \frac{1}{N_{b,u}} \right]
\approx \frac{1}{\mathbb{E} \left[ N_{b,u} \right]} 
= \frac{\lambda_b}{\lambda_u}.
\end{align}

\begin{figure}[t]
 		\centering 
 		\epsfxsize=0.48\linewidth
 		\captionsetup{width=0.48\textwidth} 
 		\includegraphics[width=8.5cm, height=6.5cm]{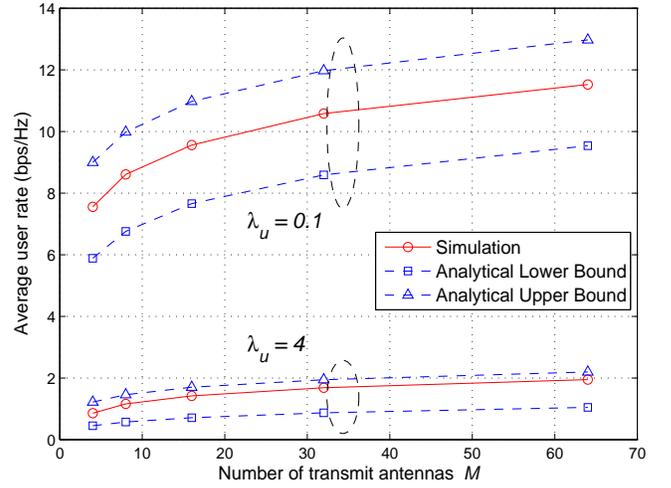}
 		\caption{Average user rate and its bounds (\ref{eq:MMIMO R UB}) and (\ref{eq:MMIMO R LB}) for M-MIMO with respect to $M$. Here, $\mu = 3.7$, $\lambda_b = 1$, and $P_T/\sigma^2 = 15$ dB.}\label{fig:ExpectRate MMIMO}
\vspace{-0.1in}
\end{figure}

\begin{figure}[t]
 		\centering 
 		\epsfxsize=0.48\linewidth
 		\captionsetup{width=0.48\textwidth} 
 		\includegraphics[width=8.5cm, height=6.5cm]{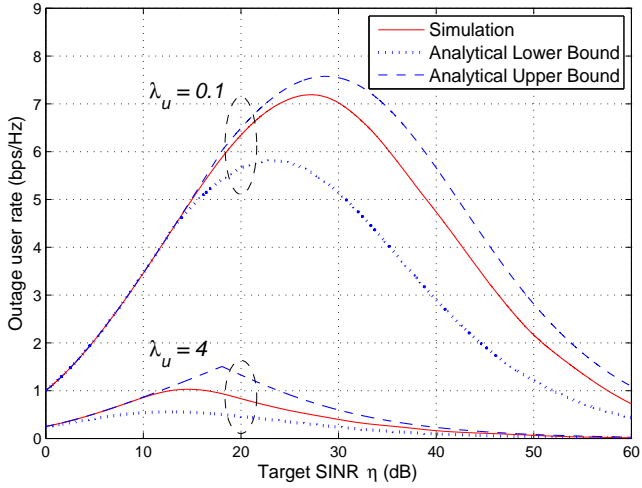}
 		\caption{Outage rate and its bounds (\ref{eq:MMIMO OR UB}) and (\ref{eq:MMIMO OR LB}) for M-MIMO with $M = 64$, $\mu = 3.7$, $\lambda_b = 1$, and $P_T/\sigma^2 = 15$ dB.}\label{fig:OutageRate MMIMO}
\vspace{-0.1in}
\end{figure}

In Figs. \ref{fig:ExpectRate MMIMO} and \ref{fig:OutageRate MMIMO}, we compare the average and outage user rates with the bounds given in (\ref{eq:MMIMO R UB})-(\ref{eq:MMIMO OR LB}). We first observe that the scaling laws of the bounds are closely matched with the true rates. The gap, however, seems to increase as the UE-BS density ratio becomes smaller. Nevertheless, the bounds are useful for further studies on the M-MIMO performance under various network models.

\subsection{Asymptotically Small UE Density}\label{subsec:asymp small UE den}

In this section, we investigate the performance of M-MIMO under the asymptotically small UE density regime. As the transmission probability $\epsilon_C$ approaches $0$, the interference also goes to $0$ since the interfering BSs are farther apart from each other. In such cases, we can ignore the interference but not the noise due to its dominant effect. The SINR can now be approximated as the SNR and expressed as 
\begin{align}\label{eq:SINR CAS Asymp}
Q_{\text{M}} 
= \text{SINR}_{b}  
\approx \text{SNR}_{b} 
& = \frac{ MP_T d_{C,b,b}^{-\mu}  \left\lVert \widehat{\mv{h}}_{C,b,b} \right\rVert^2 } { \sigma_n^2} \notag \\
& \xrightarrow{a.s.} \frac{ M^2 P_T  d_{C,b,b}^{-\mu} } { \sigma_n^2}.
\end{align}

Since the PDF of $d_{C,b,b}$ is
\begin{align}
f_{D_{C,b,b}}(x) = 2\pi \lambda_b x e^{-\pi \lambda_b x^2},
\end{align}
it is straightforward to derive the SINR distribution as follows
\begin{align}
\text{Pr} \left\{ Q_{\text{M}} \leq q \right\} 
& = \text{Pr} \left\{ d_{C,b,b} \geq \left( \frac{M^2 P_T}{q \sigma_n^2} \right)^{1/\mu} \right\} \notag \\
& = \exp \left( -\pi \lambda_b \frac{M^{4/\mu} P_T^{2/\mu}}{ \sigma_n^{4/\mu} q^{2/\mu}} \right).
\end{align}

Therefore, the average Shannon and outage rates are given as
\begin{align}\label{eq:R M Asymp}
R_{\text{M}} = \int_{0}^{\infty} \left[ 1 - \exp \left( -\pi \lambda_b \frac{M^{4/\mu} P_T^{2/\mu}}{ \sigma_n^{4/\mu} (2^t-1)^{2/\mu} } \right) \right] dt,
\end{align}

\begin{align}\label{eq:OR M Asymp}
\text{OR}_{\text{M}}(\eta) = \log_2 (1+ \eta) \left[ 1 - \exp \left( -\pi \lambda_b \frac{M^{4/\mu} P_T^{2/\mu}}{ \sigma_n^{4/\mu} \eta^{2/\mu} } \right) \right].
\end{align}

\section{Small-Cell System Analysis}\label{sec:small-cell}

Similar to M-MIMO, if the UE intensity $\lambda_u$ is not sufficiently larger than $\lambda_b$, then some DAs in the small-cell system may not be associated to any UE, and thus, do not transmit signals. Applying Lemma \ref{lemma:area PPP}, we can derive the transmission probability of a DA in our system as \cite{LeHu12}
\begin{align}
\epsilon_D \approx 1 - \left( 1+ \frac{\lambda_u}{3.5M\lambda_b} \right)^{-3.5}.
\end{align}

The transmitting DAs thus can be modeled as a homogeneous PPP $\widehat{\Phi}_{D,b}$ obtained by thinning the DA PPP $\Phi_{D,b}$ with a probability $\epsilon_D$ that DA is activated. In this case, the SINR of a typical UE $k$ can be described as
\begin{align}\label{eq:SINR Q fad}
Q_{\text{SM}} \triangleq \frac{ \big|\widehat{h}_{D,k,k} \big|^2 d_{k,k}^{-\mu} P_T}{ \sum_{k'\in \Phi_{D,b}/\{k\}}^K {\big|\widehat{h}_{k,k'}\big|^2  P_{T}}{d_{k,k'}^{-\mu}} +\sigma_n^2 }.
\end{align}

Using $\epsilon_D$, we can obtain the coverage probability of a typical user $k$ in Lemma \ref{lemma:Coverage Prob.} \cite{JoNgSu14}.
\begin{lemma}[\cite{JoNgSu14}]\label{lemma:Coverage Prob.}
Given that the transmitting DAs are modeled as PPP $\widehat{\Phi}_{D,b}$ and both desired and interfering signals obey Rayleigh distribution, the coverage probability for user $k$ is expressed as
\begin{align}\label{eq:pk DAS}
\Pr\left(Q_\text{SM} \geq q \right) 
& = \pi M\lambda_b \int_{0}^{\infty} \exp \bigg\{- \frac{\sigma_n^2}{P_T} q v^{\mu/2} \notag \\
& - \pi M\lambda_b v \left[ 1 + \epsilon_{D} q^{2/\mu} \rho \left( q^{-2/\mu}, \frac{\mu}{2} \right) \right] \bigg\} dv.
\end{align}
\end{lemma}

A direct corollary of Lemma 6 is the SIR distribution for the case $\sigma_n^2/P_T = 0$ given as follows.
\begin{corollary}
The SIR distribution of a small cell system is given as
\begin{align}\label{eq:SmC exact CDF}
F_{Q_{\text{SM}}}(q) = 1 - \frac{1}{ 1 + \epsilon_{D} q^{2/\mu} \rho \left( q^{-2/\mu}, \frac{\mu}{2} \right)}.
\end{align}
\end{corollary}

Again, we obtain the SIR CDF, average achievable rate, and outage rate bounds for the small-cell system. The differences between the following Corollaries \ref{corol:SM CDF bounds}, \ref{corol:SM LB} and \cite[Lemma 6]{NgSu15} are that the interfering cells follows a PPP with density $\epsilon_D \lambda_b$, and each DA might need to serve multiple UEs via TDMA/FDMA. We note that the expectation of the number of UEs associated with a typical DA $k$ is given as (cf. Lemma \ref{lemma:Avg N_bu})
\begin{align}
\mathbb{E} \left[ N_{k,u} \right] = \frac{\lambda_{D,u}}{\lambda_{D,b}} = \frac{\lambda_u}{M \lambda_b}. 
\end{align}

The proofs for the following results are similar to those of \cite[Lemmas 5 and 6]{NgSu15}, and thus omitted for brevity. 
\begin{corollary}\label{corol:SM CDF bounds}
The lower and upper bounds for the distribution of $Q_{\text{SM}}$ defined in (\ref{eq:SINR Q fad}) are given as
\begin{align}
F^{\text{LB}}_{Q_{\text{SM}}}(q) 
& = 1 - \mathbb{E}_{h} \left[ \frac{1}{1 + \frac{2 \epsilon_D q^{2/\mu}}{\mu |h|^{2/\mu} } \Gamma \left( \frac{2}{\mu}, \frac{|h|^2}{q} \right) } \right], \label{eq:SmC CDF LB} \\
F^{\text{UB}}_{Q_{\text{SM}}}(q) 
& = 1 - \frac{1}{1 + \epsilon_D \beta_{\text{fd}} q^{2/\mu} }, \label{eq:SmC CDF UB}
\end{align}
where $|h|^2$ is exponentially distributed. 
\end{corollary}

\begin{figure}[t]
 		\centering 
 		\epsfxsize=0.48\linewidth
 		\captionsetup{width=0.48\textwidth} 
 		\includegraphics[width=8.5cm, height=6.5cm]{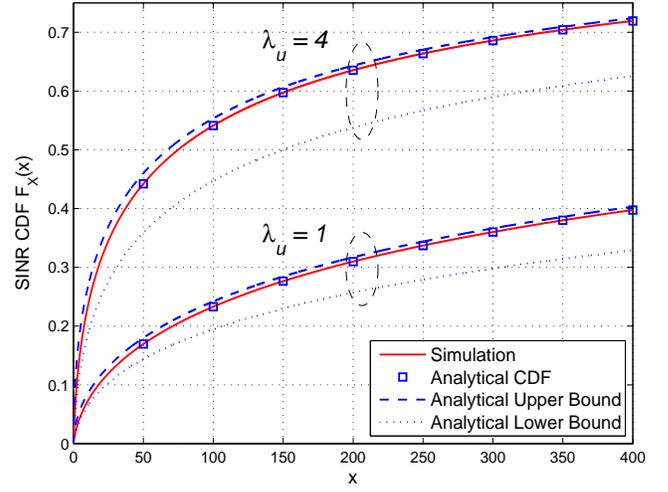}
 		\caption{Simulated SINR distribution, analytical result (\ref{eq:SmC exact CDF}), and bounds (\ref{eq:SmC CDF LB}), (\ref{eq:SmC CDF UB}) for small-cell system with $\lambda_{D,b} = M \lambda_b = 64$, $\mu = 3.7$, and $P_T/\sigma^2 = 15$ dB.}\label{fig:CDF SmallCell}
\vspace{-0.1in}
\end{figure}

\begin{corollary}\label{corol:SM LB}
The average Shannon rate $R_{\text{SM}}(\alpha)$ and the outage rate $\text{OR}_{\text{SM}}(\alpha,\eta)$ are bounded below, respectively, by 
\begin{align}
R^{\text{LB}}_{\text{SM}} 
& = \min \left\{ 1, \frac{M \lambda_b}{\lambda_u} \right\} \int_{0}^{\infty} \frac{ dt}{ 1 + \epsilon_D \beta_{\text{fd}} \left( 2^t - 1 \right)^{2/\mu} }, \label{eq:SM R LB} \\
\text{OR}^{\text{LB}}_{\text{SM}}(\eta) 
& = \min \left\{ 1, \frac{M \lambda_b}{\lambda_u} \right\} \frac{\log_2 \left(1+\eta \right)}{1 + \epsilon_D \beta_{\text{fd}} \eta^{2/\mu} }. \label{eq:SM OR LB}
\end{align}
\end{corollary}

Again, the term $\min \left\{ 1, \frac{M \lambda_b}{\lambda_u} \right\}$ is due to the fact that TDMA/FDMA is applied only when $N_{k,u} > 1$, and we have approximated
\begin{align}
\frac{1}{N_{k,u}} \approx \frac{1}{\mathbb{E} \left[ N_{k,u} \right]} = \frac{M \lambda_b}{\lambda_u}.
\end{align}

\begin{figure}[t]
 		\centering 
 		\epsfxsize=0.48\linewidth
 		\captionsetup{width=0.48\textwidth} 
 		\includegraphics[width=8.5cm, height=6.5cm]{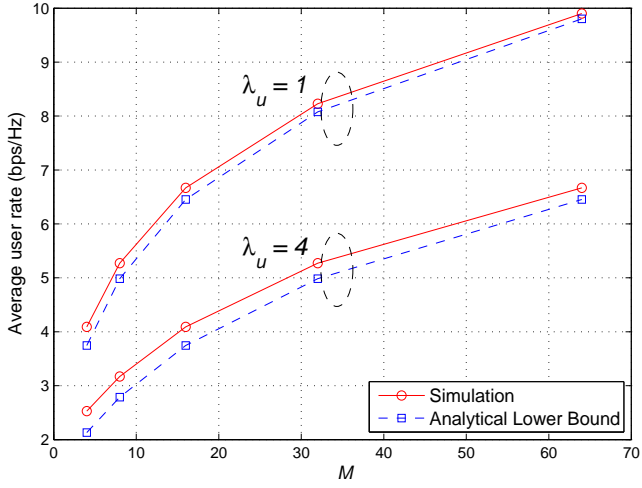}
 		\caption{Average user rate and its lower bound (\ref{eq:SM R LB}) with respect to $M$ for small-cell system. Here, $\lambda_b = 1$, $\mu = 3.7$, $P_T/\sigma^2 = 15$ dB, and $\lambda_{D,b} = M \lambda_b$.}\label{fig:ExpectRate SmCell}
\vspace{-0.1in}
\end{figure}

\begin{figure}[t]
 		\centering 
 		\epsfxsize=0.48\linewidth
 		\captionsetup{width=0.48\textwidth} 
 		\includegraphics[width=8.5cm, height=6.5cm]{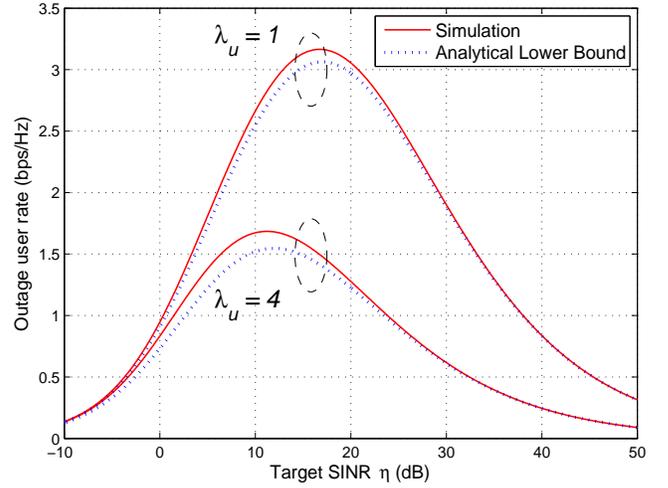}
 		\caption{Outage rate and its lower bound (\ref{eq:SM OR LB}) for small-cell system with $\lambda_{D,b} = M \lambda_b = 16$, $\mu = 3.7$, and $P_T/\sigma^2 = 15$ dB.}\label{fig:OutageRate SmCell}
\vspace{-0.1in}
\end{figure}

In Fig. \ref{fig:CDF SmallCell}, we compare the SINR distribution, analytical result (\ref{eq:SmC exact CDF}), and bounds (\ref{eq:SmC CDF LB}), (\ref{eq:SmC CDF UB}) for two small-cell systems. It is observed that the upper-bound (\ref{eq:SmC CDF UB}), although simple, is quite tight when compared with the lower-bound (\ref{eq:SmC CDF LB}). Figs. \ref{fig:ExpectRate SmCell} and \ref{fig:OutageRate SmCell} illustrate the average and outage user rate for various small-cell systems. We also show the analytical rate lower bounds given in Corollary \ref{corol:SM LB} for comparison. In contrast to the M-MIMO case, the bounds for small-cell systems are quite tight. Note that we do not derive and present the analytical upper bounds for the rates here since they are complicated and of limited use for subsequent analyses. 

\subsection{Asymptotically Small UE Density}\label{subsec:asymp small UE den SM}

Under the asymptotically small UE density regime, the interference is negligible and the SINR of small-cell system can be expressed as 
\begin{align}\label{eq:SINR SM Asymp}
Q_{\text{SM}} = \text{SINR}_{k}  
\approx \text{SNR}_{k} = \frac{ \big| \widehat{h}_{D,k,k} \big|^2 d_{k,k}^{-\mu} P_T } { \sigma_n^2}.
\end{align}

Since the PDF of $d_{k,k}$ is
\begin{align}
f_{D_{k,k}}(x) = 2\pi M \lambda_b x e^{-\pi M \lambda_b x^2},
\end{align}
it is straightforward to derive the SINR distribution as follows
\begin{align}\label{eq:OP SM Asymp}
\text{Pr} \left\{ Q_{\text{SM}} \leq q \right\} 
& = \mathbb{E}_h \left[ \exp \left( -\pi M \lambda_b \frac{ P_T^{2/\mu} |h|^{2/\mu}}{ \sigma_n^{4/\mu} q^{2/\mu}} \right) \right],
\end{align}
where the random variable $h \sim \mathcal{CN}(0,1)$.

Therefore, the average Shannon and outage rate are
\begin{align}\label{eq:R SM Asymp}
R_{\text{SM}} = \int_{0}^{\infty} \left( 1 - \mathbb{E}_h \left[ \exp \left( -\pi M \lambda_b \frac{ P_T^{2/\mu} |h|^{2/\mu}}{ \sigma_n^{4/\mu} (2^t-1)^{2/\mu}} \right) \right] \right) dt,
\end{align}

\begin{align}\label{eq:OR SM Asymp}
& \text{OR}_{\text{SM}}(\eta) \notag \\
& = \log_2 (1+ \eta) \left( 1 - \mathbb{E}_h \left[ \exp \left( -\pi M \lambda_b \frac{ P_T^{2/\mu} |h|^{2/\mu}}{ \sigma_n^{4/\mu} \eta^{2/\mu}} \right) \right] \right).
\end{align}

To derive the bounds for (\ref{eq:OP SM Asymp})-(\ref{eq:OR SM Asymp}), we note that the function $\exp(-\alpha x^{\beta})$ with $\beta\leq 1$ is convex in $x >0$. Therefore, applying Jensen's inequality, we obtain
\begin{align}
F_{Q_{\text{SM}}}(q) 
& \geq  \exp \left( -\pi M \lambda_b \frac{ P_T^{2/\mu} \mathbb{E}_h \left[ |h|^{2/\mu} \right] }{ \sigma_n^{4/\mu} q^{2/\mu}} \right) \notag \\
& = \exp \left( -\pi M \lambda_b \frac{ P_T^{2/\mu} \Gamma(1+\frac{2}{\mu}) }{ \sigma_n^{4/\mu} q^{2/\mu}} \right)
\triangleq F^{\text{LB}}_{Q_{\text{SM}}}(q).
\end{align}

The upper bounds for $R_{\text{SM}}$ and $\text{OR}_{\text{SM}}(\eta)$ are therefore given as
\begin{align}\label{eq:R SM Asymp UB}
R_{\text{SM}}^{\text{UB}} = \int_{0}^{\infty} \left( 1 - \exp \left( -\pi M \lambda_b \frac{ P_T^{2/\mu} \Gamma(1+\frac{2}{\mu}) }{ \sigma_n^{4/\mu} (2^t-1)^{2/\mu}} \right) \right) dt,
\end{align}

\begin{align}\label{eq:OR SM Asymp UB}
& \text{OR}_{\text{SM}}^{\text{UB}}(\eta) \notag \\
& = \log_2 (1+ \eta) \left( 1 - \exp \left( -\pi M \lambda_b \frac{ P_T^{2/\mu} \Gamma(1+\frac{2}{\mu}) }{ \sigma_n^{4/\mu} q^{2/\mu}} \right) \right).
\end{align}

\section{Rate Comparison for M-MIMO and Small-Cell Systems}\label{sec:comparison}

M-MIMO and small-cell represent two densification approaches in 5G communication systems. The lack of a performance comparison between them is even more surprising given the fact that each approach has attracted significant attention in the literature. In this section, we exploit the newly obtained results in Sections \ref{sec:M-MIMO CAS} and \ref{sec:small-cell} to draw several observations on M-MIMO and small-cell networks, revealing which approach is better and under which setup. Here, the metric is the outage rate for simple arguments, but our observations can be readily extended to the average Shannon rate. The key parameter here is the UE density $\lambda_u$. 

\subsection{Very Large UE Density: $\lambda_u \geq M \lambda_b$}\label{subsec:compare 1}

In this case, the lower bounds of the outage rate are reduced to
\begin{align}\label{eq:MMIMO OR LB 1}
\text{OR}^{\text{LB}}_{\text{M}}(\eta) = \frac{\lambda_b}{\lambda_u} \frac{\log_2 \left(1+\eta \right)}{1 + \epsilon_C \beta_{\text{n-fd}} M^{-2/\mu} \eta^{2/\mu} },
\end{align}

\begin{align}\label{eq:SM OR LB 1}
\text{OR}^{\text{LB}}_{\text{SM}}(\eta) = \frac{M \lambda_b}{\lambda_u} \frac{\log_2 \left(1+\eta \right)}{1 + \epsilon_D \beta_{\text{fd}} \eta^{2/\mu} },
\end{align}
where $\epsilon_C \approx 1 - \left( 1+ \frac{\lambda_u}{3.5\lambda_b} \right)^{-3.5} \approx 1$. Simple manipulation shows that $\text{OR}^{\text{LB}}_{\text{SM}}(\eta) \geq \text{OR}^{\text{LB}}_{\text{M}}(\eta)$ iff 
\begin{align}\label{eq: comp 1}
M + M^{1-2/\mu} \epsilon_C \beta_{\text{n-fd}} \eta^{2/\mu} \geq 1 + \epsilon_D \beta_{\text{fd}} \eta^{2/\mu}.
\end{align}

As a loose approximation, (\ref{eq: comp 1}) holds if $M^{1-2/\mu} \epsilon_C \beta_{\text{n-fd}} \geq  \epsilon_D \beta_{\text{fd}}$, which is true since $\beta_{\text{n-fd}} > \beta_{\text{fd}}$, $\epsilon_C \approx 1 \geq \epsilon_D$, and $M$ is large. We thus observe that a small-cell system outperforms a M-MIMO counterpart when the UE density is very large. This is due to the effect of multiplexing large number of users under TDMA/FDMA, which diminishes the interference-mitigation benefit of M-MIMO.

\subsection{Intermediate UE density: $\lambda_b < \lambda_u < M \lambda_b$}\label{subsec:compare 2}

Since $\frac{\lambda_u}{3.5 M \lambda_b}$ is small, we have
\begin{align}
\epsilon_D \approx 1 - \left( 1+ \frac{\lambda_u}{3.5 M \lambda_b} \right)^{-3.5} \approx \frac{\lambda_u}{M \lambda_b}.
\end{align}

Based on (\ref{eq:MMIMO OR LB}) and (\ref{eq:SM OR LB}), we can express the outage rates as
\begin{align}\label{eq:MMIMO OR LB 3}
\text{OR}^{\text{LB}}_{\text{M}}(\eta) = \frac{\lambda_b}{\lambda_u} \frac{\log_2 \left(1+\eta \right)}{1 + \epsilon_C \beta_{\text{n-fd}} M^{-2/\mu} \eta^{2/\mu} },
\end{align}

\begin{align}\label{eq:SM OR LB 3}
\text{OR}^{\text{LB}}_{\text{SM}}(\eta) = \frac{\log_2 \left(1+\eta \right)}{1 + \epsilon_D \beta_{\text{fd}} \eta^{2/\mu} }.
\end{align}

The condition for $\text{OR}^{\text{LB}}_{\text{SM}}(\eta) \geq \text{OR}^{\text{LB}}_{\text{M}}(\eta)$ is
\begin{align}\label{eq:compare 1}
1 +  \frac{1}{M^{2/\mu}} \epsilon_C \beta_{\text{n-fd}}  \eta^{2/\mu}  \geq  \frac{\lambda_b}{\lambda_u} + \frac{1}{M} \beta_{\text{fd}} \eta^{2/\mu}.
\end{align}

We note that (\ref{eq:compare 1}) is true since $1 \geq \frac{\lambda_b}{\lambda_u}$ and $M \geq M^{2/\mu}$. Therefore, s small-cell system outperforms a M-MIMO counterparts when the UE density is moderate. Similar to Section \ref{subsec:compare 1}, the reason is again due to the effect of multiplexing a large number of users under TDMA/FDMA M-MIMO. 

\begin{remark}
In Sections \ref{subsec:compare 1} and \ref{subsec:compare 2}, a stronger and stricter result which states that $\text{OR}^{\text{LB}}_{\text{SM}}(\eta) \geq \text{OR}^{\text{UB}}_{\text{M}}(\eta)$ is also provable using the same arguments. However, we have used the lower bounds $\text{OR}^{\text{LB}}_{\text{M}}(\eta)$ instead due to the symmetry and for the ease of explanation. 
\end{remark}

\subsection{Asymptotically Small UE Density: $\lambda_u \ll \lambda_b$}\label{subsec:compare 3}

Assuming a small UE density, it is difficult to compare M-MIMO and small-cell systems since the M-MIMO lower bounds are not close to the exact distribution or rates. In this subsection, we consider the asymptotic regime where $\lambda_u$ is very small compared to $\lambda_b$ to reveal some insights for the small UE density case. Note that
\begin{align}\label{eq:asymp 1}
\exp \left( - \frac{ \pi \lambda_b P_T^{2/\mu}}{ \sigma_n^{4/\mu} q^{2/\mu}} M^{4/\mu} \right) 
\leq \exp \left( - \frac{ \pi \lambda_b P_T^{2/\mu}  }{ \sigma_n^{4/\mu} q^{2/\mu}} M \Gamma \Big( 1+\frac{2}{\mu} \Big) \right),
\end{align}
when $\mu < 4$ and $M$ is large enough, i.e., $M^{\frac{4}{\mu}-1} > \Gamma \Big( 1+\frac{2}{\mu} \Big)$. From (\ref{eq:R M Asymp}), (\ref{eq:OR M Asymp}), (\ref{eq:R SM Asymp UB}), and (\ref{eq:OR SM Asymp UB}), it is straightforward to prove that $R_{\text{M}} \geq R^{\text{UB}}_{\text{SM}}$ and
$\text{OR}_{\text{M}}(\eta) \geq \text{OR}^{\text{UB}}_{\text{SM}}(\eta)$. 

\begin{figure}[t]
 		\centering 
 		\epsfxsize=0.48\linewidth
 		\captionsetup{width=0.48\textwidth} 
 		\includegraphics[width=8.5cm, height=6.5cm]{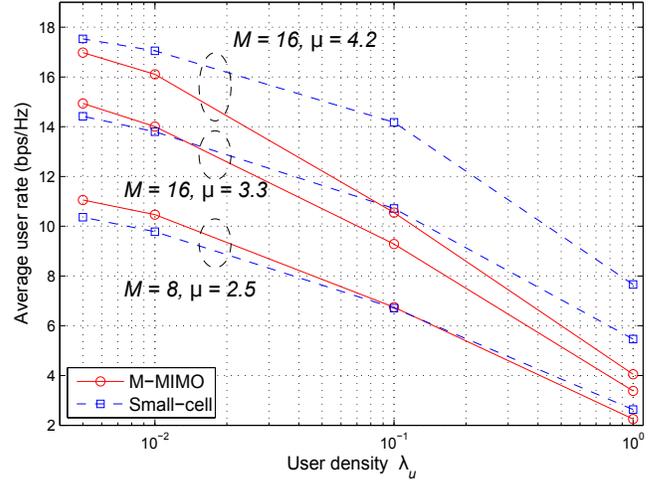}
 		\caption{Comparison of the achievable user rate between M-MIMO and small-cell systems with $\lambda_{b} = 1$.}\label{fig:RateCmpr MMIMO SmCell}
\vspace{-0.1in}
\end{figure}

The above result essentially asserts that M-MIMO outperforms small-cell densification, albeit when the UE density is asymptotically small and $\mu < 4$. Combined with observations from Sections \ref{subsec:compare 1} and \ref{subsec:compare 2}, we note that there exists a threshold with smaller UE density than which we should employ M-MIMO and larger UE density than which small-cell densification is more preferable, in terms of spectral efficiency. As an example, we compare the achievable user rates of M-MIMO and small-cell systems in Fig. \ref{fig:RateCmpr MMIMO SmCell}. Note that as the path-loss exponent $\mu$ approaches $4$, $\lambda_u$ needs to be smaller so that M-MIMO rate can outperform small-cell counterpart. When $\mu \geq 4$, the M-MIMO rate is always worse than that of small-cell, as suggested by (\ref{eq:asymp 1}).

We would like to highlight here that our study have not considered the deployment cost, the signaling overhead, the handover implementation, etc. We have also not considered multiuser beamforming in M-MIMO. Due to the space constraint, we leave such interesting and important investigations as our future works.

Note that we can consider small-cell and M-MIMO systems as two extrema of a balancing problem, where the number of BS antennas and BS density can vary while their product always equals to $M\lambda_b$. In other words, the problem is to distribute the antennas from a pool with density $M\lambda_b$ so that the resulting rate performance is optimal. The analysis of this problem is partially covered by Sections \ref{sec:M-MIMO CAS} and \ref{sec:small-cell}. The above arguments suggest that the optimal setup should be a moderate M-MIMO system with sufficient beamforming gain and a sufficiently dense BS deployment.  

\section{Energy Efficiency Comparison for M-MIMO and Small-Cell Systems}\label{sec:comparison EE}

Apart from spectral efficiency, energy efficiency (EE) is also an important factor for evaluating future communications networks due to several environmental concerns. In this section, we compare M-MIMO and small-cell systems using EE as the metric. In the literature, a conventional definition for EE is
\begin{align}\label{eq:con EE}
\text{EE} \triangleq  \frac{ R_{\text{AP/BS}} }{P_{\text{AP/BS}}},
\end{align}
where $P_{\text{AP/BS}}$ and $R_{\text{AP/BS}}$ are the transmit power and sum-rate of the corresponding AP/BS to all users, respectively. However, for ease of analysis, we consider the following EE definition based on the outage rate
\begin{align}
\text{EE}(\eta) \triangleq  \frac{  \text{OR}_{\text{AP/BS}} (\eta) }{P_{\text{AP/BS}}}.
\end{align}

We can interpret $\text{EE}(\eta)$ as the energy efficiency of the AP/BS given that the user SINR satisfies the constraint SINR $\geq \eta$. Observations for EE based on (\ref{eq:con EE}) can be similarly obtained. Note that due to the transmission probability, the expected transmission power for each M-MIMO BS and small-cell AP are given as $\epsilon_C M P_T$ and $\epsilon_D P_T$, respectively. We first consider the moderate to large UE density regime where $\lambda_b < \lambda_u$ and obtain 
\begin{align}
\text{EE}^{\text{LB}}_{\text{SM}}(\eta) 
= \frac{ \text{OR}^{\text{LB}}_{\text{SM}}(\eta) }{\epsilon_D P_T} 
> \frac{ \text{OR}^{\text{LB}}_{\text{M}}(\eta) }{\epsilon_D P_T} 
\gg \frac{ \text{OR}^{\text{LB}}_{\text{M}}(\eta) }{\epsilon_C M P_T}
= \text{EE}^{\text{LB}}_{\text{M}}(\eta), 
\end{align} 
since $\epsilon_D < \epsilon_C$, $M \gg 1$, and $\text{OR}^{\text{LB}}_{\text{SM}}(\eta) > \text{OR}^{\text{LB}}_{\text{M}}(\eta)$ from Sections \ref{subsec:compare 1} and \ref{subsec:compare 2}.

Furthermore, under the asymptotically small UE density regime with $\lambda_u \ll \lambda_b$, we have $\epsilon_C \approx \frac{\lambda_u}{\lambda_b}$ and $\epsilon_D \approx \frac{\lambda_u}{M \lambda_b}$. Therefore, $\epsilon_C \gg \epsilon_D$. From (\ref{eq:R M Asymp}), (\ref{eq:OR M Asymp}), (\ref{eq:R SM Asymp}), and (\ref{eq:OR SM Asymp}), it is straightforward to show that $\text{EE}_{\text{SM}} > \text{EE}_{\text{M}}$ and $\text{EE}_{\text{SM}}(\eta) > \text{EE}_{\text{M}}(\eta)$ by noting that the exponential function approaches $0$ as $M$ grows large. 

The above result reveals that small-cell systems are more energy-efficient than M-MIMO counterparts with large, moderate, or asymptotically small UE densities. Strictly speaking, such results do not cover the small UE density regime. However, we conjecture that small cells outperforms M-MIMO in terms of EE under all cases. This conclusion is likely to hold even with multiuser beamforming/precoding applied at each M-MIMO BS.

From Sections \ref{sec:comparison} and \ref{sec:comparison EE}, we observe another important aspect of the trade-off between M-MIMO and small-cell densification. Particularly, even through M-MIMO have a higher spectral efficiency as under small UE density regime, small-cell densification might still be a preferable option for certain communication systems due to its higher EE performance.   

\section{Conclusions}\label{sec:conclusions}

In this paper, we have compared the spectral and energy efficiencies of massive MIMO and small-cell systems. Particularly, we have derived SIR distribution bounds for both systems, based on which the average Shannon and outage rate bounds are obtained. We have also analyzed the performance of M-MIMO and small-cell under asymptotically small UE density regime, which represents UE-sparse networks. The M-MIMO and small-cell systems were then compared in terms of spectral and energy efficiencies. For the rate performance, we observe that M-MIMO surpasses small-cell densification when the UE density is small compared with BS/AP density and number of antennas, and vice versa. However, small-cell network yields better energy efficiency than M-MIMO counterpart under all cases. The results of this paper are useful for the optimal design of practical 5G networks and other 5G system performance analyses.   

\appendices

\section{Proof of Corollaries \ref{corol:MMIMO CDF SINR} and \ref{corol:MMIMO CDF bounds}}\label{appen:MMIMO CDF}

\subsection{Proof of Corollary \ref{corol:MMIMO CDF SINR}}\label{appen:MMIMO CDF SINR}

Given that $d_{C,b,b} = x$. The Laplace transform is given as \cite[(26)]{NgSu15}
\begin{align}
& \mathcal{L}_{Q_\text{M}^{-1}} \left(s|d_{C,b,b}=x \right) 
= e^{- \frac{s}{M} x^{\mu} \delta} \exp \bigg\{ - \pi \widehat{\lambda}_{C,b} x^2 (1-e^{-s/M}) \notag \\
& \qquad \qquad  - \pi \widehat{\lambda}_{C,b} \left( \frac{s}{M} \right)^{2/\mu} x^{2} \gamma \left( 1- \frac{2}{\mu},\frac{s}{M} \right)  \bigg\},
\end{align}
where $\delta = \frac{\sigma_n^2}{MP_T}$. Now we observe that $d_{C,b,b}$ is the distance from an arbitrary origin to the nearest point of a PPP with density $\lambda_b$. The PDF of $d_{C,b,b}$ is expressed as follows
\begin{align}
f_{D_{C,b,b}}(x) = 2\pi \lambda_b x e^{-\pi \lambda_b x^2}. 
\end{align}

The Laplace transform of $\text{SINR}_b^{-1}$ is thus obtained as
\begin{align}
& \mathcal{L}_{Q_\text{M}^{-1}} \left(s \right) \notag \\
& = \int_{0}^{\infty} \mathcal{L}_{\text{SINR}^{-1}} \left(s|d_{C,b,b}=x \right) f_{D_{C,b,b}}(x) dx \notag \\
& = 2\pi \lambda_b \int_{0}^{\infty} x \exp \bigg\{ - \frac{s}{M} x^{\mu} \delta - \pi \widehat{\lambda}_{C,b} x^2 \bigg[ 1 + \frac{\lambda_b}{\widehat{\lambda}_{C,b}} \notag \\
& \qquad \qquad - e^{-s/M} + \left( \frac{s}{M} \right)^{2/\mu} \gamma\left( 1- \frac{2}{\mu}, \frac{s}{M} \right)  \bigg]   \bigg\} dx \notag \\
& = \pi \lambda_b \int_{0}^{\infty}  \exp \bigg\{ -\frac{s y^{\mu/2} \sigma_n^2}{M^2 P_T} - \pi \widehat{\lambda}_{C,b} y \bigg[ 1 + \frac{1}{\epsilon_C} \notag \\
& \qquad \qquad - e^{-s/M} + \left( \frac{s}{M} \right)^{2/\mu} \gamma\left( 1- \frac{2}{\mu}, \frac{s}{M} \right)  \bigg]   \bigg\} dy.
\end{align}

This concludes the proof of Corollary \ref{corol:MMIMO CDF SINR}.

\subsection{Proof of Corollary \ref{corol:MMIMO CDF bounds}}\label{appen:MMIMO CDF bounds}

For ease of presentation, we will consider $Q_\text{M}/M$. The lower bound is given as
\begin{align}
F^{\text{LB}}_{Q_\text{M}/M}(q) 
& = \int_{0}^{\infty} F^{\text{LB}}_{Q_M/M}(q|d_{C,b,b}=x) f_{D_{C,b,b}}(x) dx  \notag \\
& = \int_{0}^{\infty} \left(  1 - e^{- \widehat{\lambda}_{C,b} \pi \left( q^{2/\mu}-1 \right) x^2} \right) 2 \pi \lambda_b x e^{-\pi \lambda_b x^2} dx \notag \\
& = 1 - \frac{1}{ \epsilon_C q^{2/\mu} + 1 - \epsilon_C}.
\end{align}

Now assume that $d_{C,b,b} = x$, $q \geq 1$, and $\xi \geq 1$. Similar to \cite[(39)]{NgSu15}, the distribution for massive MIMO case can be bounded as 
\begin{align}
& F_{Q_\text{M}/M}(q|d_{C,b,b}=x)
\leq 1 - e^{- \widehat{\lambda}_{C,b} \pi \left( (\xi q)^{2/\mu}-1 \right) x^2} \notag \\
& \qquad  + e^{ - \widehat{\lambda}_{C,b} \pi \left( (\xi q)^{2/\mu}-1 \right) x^2} \frac{2\pi \widehat{\lambda}_{C,b} x^2 \xi^{2/\mu-1} q^{2/\mu}}{\mu-2},
\end{align}

We therefore obtain
\begin{align}
& F_{Q_\text{M}/M}(q) \notag \\
& \leq 2\pi \lambda_b \int_{0}^{\infty} \left(  1 - e^{- \widehat{\lambda}_{C,b} \pi \left( (\xi q)^{2/\mu}-1 \right) x^2} \right) x e^{-\pi \lambda_b x^2} dx \notag \\
& + \frac{ (2\pi \lambda_b)^2 \epsilon_C \xi^{2/\mu-1} q^{2/\mu}}{\mu-2} \int_{0}^{\infty} x^3 e^{-\pi \lambda_b \left( \epsilon_C (\xi q)^{2/\mu} + 1 - \epsilon_C \right) x^2} dx  \notag \\
& = 1 - \frac{1}{\epsilon_C (\xi q)^{2/\mu} + 1 - \epsilon_C} \notag \\
& \qquad \qquad + \frac{2 \epsilon_C \xi^{2/\mu-1} q^{2/\mu}}{\mu-2} \frac{1}{ \left( \epsilon_C (\xi q)^{2/\mu} + 1 - \epsilon_C \right)^2} \label{eq:appen 4} \\
& \overset{(a)}{=} 1 - \frac{1}{ \widehat{\xi}^{2/\mu} \left( \epsilon_C q^{2/\mu} + 1 - \epsilon_C \right)} \notag \\
& \qquad \qquad + \frac{2 \xi^{2/\mu-1} \epsilon_C  q^{2/\mu}}{\mu-2} \frac{1}{ \widehat{\xi}^{4/\mu} \left( \epsilon_C q^{2/\mu} + 1 - \epsilon_C \right)^2} \notag \\
& \overset{(b)}{\leq} 1 - \frac{(\mu-2)\widehat{\xi}-2}{(\mu-2)\widehat{\xi}^{1+2/\mu}} \frac{1}{\epsilon_C q^{2/\mu} + 1 - \epsilon_C},
\end{align}
where in $(a)$ we have defined 
\begin{align}
\widehat{\xi} \triangleq \frac{\epsilon_C (\xi q)^{2/\mu} + 1 - \epsilon_C}{\epsilon_C q^{2/\mu} + 1 - \epsilon_C};
\end{align}
and $(b)$ is due to the fact that $\xi^{2/\mu-1} \leq \widehat{\xi}^{2/\mu-1} $ with $\mu>2$ and $0 \leq \epsilon_C \leq 1$. Note that $\widehat{\xi}$ has the same support as $\xi$, i.e., $\widehat{\xi} \in [1,\infty)$. The next step is to find the maximum of the function
\begin{align}
f\left( \widehat{\xi} \right) = \frac{(\mu-2)\widehat{\xi}-2}{(\mu-2)\widehat{\xi}^{1+2/\mu}}
\end{align}
given that $\widehat{\xi} \geq 1$. The optimal $\widehat{\xi}^*$ is $\frac{\mu+2}{\mu-2}$. This value is a guidance for our subsequent analysis. 

Specifically, from (\ref{eq:appen 4}),  we will show that 
\begin{align}\label{eq:appen 5}
& 1 - \frac{1}{1 + \epsilon_C \beta_{\text{n-fd}} q^{2/\mu}} 
\geq 1 - \frac{1}{\epsilon_C (\xi_0 q)^{2/\mu} + 1 - \epsilon_C} \notag \\
& \qquad \qquad + \frac{2 \epsilon_C \xi_0^{2/\mu-1} q^{2/\mu}}{\mu-2} \frac{1}{ \left( \epsilon_C (\xi_0 q)^{2/\mu} + 1 - \epsilon_C \right)^2} ,
\end{align}
where $\xi_0 = \frac{\mu+2}{\mu-2}$. After some manipulations, it is observed that (\ref{eq:appen 5}) holds true if and only if
\begin{align}\label{eq:appen 6}
& \frac{2}{\mu-2} q^{2/\mu} \xi_0^{2/\mu-1} + \frac{2}{\mu-2} \xi_0^{2/\mu-1} q^{4/\mu} \epsilon_C \beta_{\text{n-fd}} \notag \\
& \qquad \leq \epsilon_C \xi_0^{2/\mu} \left( \beta_{\text{n-fd}} - \xi_0^{2/\mu} \right) q^{4/\mu} + 1 - \epsilon_C + \epsilon_C (\xi_0 q)^{2/\mu} \notag \\
& \qquad \qquad + (1 - \epsilon_C) \left( \beta_{\text{n-fd}} - \xi_0^{2/\mu} \right) q^{2/\mu}.
\end{align}

Now note that with $\beta_{\text{n-fd}} = \frac{(\mu+2)^{2/\mu+1}}{\mu (\mu-2)^{2/\mu}}$ and $\xi_0 = \frac{\mu+2}{\mu-2}$, we have
\begin{align}
\beta_{\text{n-fd}} - \xi_0^{2/\mu} = \frac{2}{\mu-2} \beta_{\text{n-fd}} \frac{1}{\xi_0} = \frac{2}{\mu+2} \beta_{\text{n-fd}},
\end{align}
which leads to
\begin{align}\label{eq:appen 7}
\epsilon_C \xi_0^{2/\mu} \left( \beta_{\text{n-fd}} - \xi_0^{2/\mu} \right) q^{4/\mu} = \frac{2}{\mu-2} \xi_0^{2/\mu-1} q^{4/\mu} \epsilon_C \beta_{\text{n-fd}}.
\end{align}

Furthermore, we obtain
\begin{align}\label{eq:appen 8}
& \epsilon_C (\xi_0 q)^{2/\mu} + (1 - \epsilon_C) \left( \beta_{\text{n-fd}} - \xi_0^{2/\mu} \right) q^{2/\mu} - \frac{2}{\mu-2} q^{2/\mu} \xi_0^{2/\mu-1} \notag \\
& = q^{2/\mu} \frac{(\mu+2)^{2/\mu-1}}{(\mu-2)^{2/\mu}} \left( \epsilon_C(\mu+2) +  (1 - \epsilon_C) \frac{2(\mu+2)}{\mu} - 2 \right) \notag \\
& = q^{2/\mu} \frac{(\mu+2)^{2/\mu-1}}{(\mu-2)^{2/\mu}} \left( \epsilon_C \mu +  (1 - \epsilon_C) \frac{4}{\mu} \right) \geq 0.
\end{align}

Combining (\ref{eq:appen 7}) and (\ref{eq:appen 8}) we thus showed that (\ref{eq:appen 6}) holds true. Therefore, for $q \geq 1$, we conclude that
\begin{align}
F_{Q_\text{M}/M}(q) \leq 1 - \frac{1}{1 + \epsilon_C \beta_{\text{n-fd}} q^{2/\mu}} = F^{\text{UB}}_{Q_\text{M}/M}(q).
\end{align}

Similar to \cite[Lemma 2]{NgSu15}, we can show that 
\begin{align}\label{eq:appen 2}
F_{Q_\text{M}/M}(q) < 1 - \frac{1}{1 + \epsilon_C \beta_{\text{n-fd}} q^{2/\mu}}
\end{align}
as $q\to 0$ or $q<1$ but $q \to 1$. Intensive simulations show that (\ref{eq:appen 2}) holds true for all intermediate values $0< q < 1$ as well.

This concludes the proof of Corollary \ref{corol:MMIMO CDF bounds}.

\section{Proof of Lemma \ref{lemma:Avg N_bu}}\label{appen:Avg N_bu}

Applying Lemma \ref{lemma:area PPP}, the PDF of $S$, the area of a typical Voronoi cell formed from $\Phi_{C,b}$, can be given as \cite[(12)]{FeNe07}
\begin{align}
f_S(x) = \frac{3.5^{3.5}}{\Gamma(3.5)} \lambda_b^{3.5} x^{2.5} e^{-3.5 \lambda_b x}.
\end{align}

Since the UEs are distributed according to a homogeneous PPP with density $\lambda_u$, the probability of the number of UEs in an area $x$ is given by
\begin{align}
\text{Pr} \{ N_{u} = n | \text{area} = x \} = e^{-\lambda_u x} \frac{\lambda_u^n x^n}{n!}.
\end{align}

The distribution of the number of UEs is thus equal to
\begin{align}\label{eq:Pr N=n}
& \text{Pr} \{ N_{b,u} = n \} 
= \int_{0}^{\infty} \text{Pr} \{ N_{b,u} = n | \text{area} = x \} f_S(x) dx \notag \\
& = \frac{3.5^{3.5}}{\Gamma(3.5)} \lambda_b^{3.5} \frac{\lambda_u^n}{n!} \int_{0}^{\infty} x^{n+2.5} e^{-(\lambda_u + 3.5 \lambda_b) x} dx 
\end{align}
Note that by using \cite[(3.371)]{GrRy07}, we can express the distribution of the number of UEs as
\begin{align}\label{eq:prob N_b} 
& \text{Pr} \{ N_{b,u} = n \} \notag \\
& \qquad = \frac{(2n+5)!!}{n!} \frac{3.5^{3.5}}{15\times 2^n} \lambda_b^{3.5} \lambda_u^n \left( \frac{1}{\lambda_u + 3.5 \lambda_b} \right)^{n+3.5}.
\end{align}
This expression, however, is difficult to manipulate. In the following derivation, we use (\ref{eq:Pr N=n}) instead. The expectation of $N_{b,u}$ is given as
\begin{align}\label{eq:ext K}
& \mathbb{E} \left[ N_{b,u} \right] 
= \sum_{n=1}^{\infty} n \text{Pr} \{ N_{b,u} = n \} \notag \\
& \qquad = \frac{3.5^{3.5}}{\Gamma(3.5)} \lambda_b^{3.5}  \int_{0}^{\infty} \left( \sum_{n=1}^{\infty} n \frac{\lambda_u^n x^n}{n!} \right) n^{2.5} e^{-(\lambda_u + 3.5 \lambda_b) x} dx \notag \\
& \qquad \overset{(a)}{=} \frac{3.5^{3.5}}{\Gamma(3.5)} \lambda_b^{3.5} \lambda_u \int_{0}^{\infty} e^{-3.5 \lambda_bx} x^{3.5} dx \overset{(b)}{=} \frac{\lambda_u}{\lambda_b},
\end{align}
where $(a)$ comes from the fact that $\sum_{n=1}^{\infty} n \frac{x^n}{n!} = e^{x} x$ and $(b)$ is due to \cite[(3.371)]{GrRy07}. This concludes the proof of Lemma \ref{lemma:Avg N_bu}.



\balance
\end{document}